\documentclass{PoS}

\title{From lepton interactions to hadron and nuclear ones at high multiplicity}

\ShortTitle{From lepton interactions to hadron and nuclear ones at high multiplicity}

\author{V. Dunin$^a$, \speaker{E. Kokoulina}$^{a}$, 
M.~Nevmerzhitsky$^b$, V. Nikitin$^a$, Yu.~Petukhov$^a$,
V.~Riadovikov$^c$,
I.~Roufanov$^a$, V.~Volkov$^d$ and A. Vorobiev$^c$\\
 \llap{$^a$}Joint Institute for Nuclear Research,
  Joliot-Curie 6, Dubna, Moscow region, 141980 Russia\\
  \llap{$^b$} Institute for Applied Physics,
 Akademicheskaya 16, Minsk 220072, Belarus\\
 \llap{$^c$} Institute of High Energy Physics, 
 Sq. Nauki 1, Protvino, Moscow region, 142281 Russia\\
 \llap{$^d$} M.V. Lomonosov MSU, SINP MSU,
 Leninskie gory, Moscow 119991, Russia\\ 
  E-mail:  \email{dunin@gmail.com}, \email{kokoulina@jinr.ru}, 
\email{nevmerzhMN@gmail.com}, \email{nikitin@sunse.jinr.ru},
\email{Yuri.Petukhov@ihep.ru},
\email{riadovikov@ihep.ru}, \email{roufanov@gmail.com},
\email{volkov@mail.desy.de},
\email{vorobiev@ihep.ru}}

\abstract{Multiplicity data up to 200 GeV in $e^+e^-$ annihilation are described well by the two-stage model based on pQCD and suggested the phenomenological scheme of hadronization. This model confirms the fragmentation mechanism of hadronization (in vacuum). It  allows to estimate mean multiplicity at 500 GeV and 1 TeV. 
Gluon dominance model is the modification of this model for the description of hadronic interactions. It  was realised by an inclusion of gluons. It demonstrates very strong evidence of the recombination mechanism of hadronization. In this case, the mean multiplicity of hadrons formed from a single gluon grows with energy and it exceeds the corresponding values for lepton interactions. At the same time, the region of high multiplicity is stipulated for splitting of active gluons. The excess of soft photon yield is experimentally confirmed at Nuclotron (JINR) in the interactions of the 3.5A GeV/$c$ deuteron and lithium beams with the carbon target.}

\FullConference{38th International Conference on High Energy Physics\\
                  3-10 August 2016\\
                 Chicago, USA}

\begin{document}

\section{Lepton interactions}
In accordance with present understanding, the multi-particle production in $e^+e^-$ annihilation occurs through two stages: development of a quark-gluon ($qg$) cascade  and hadronisation: $e^+e^-\to q \overline q \to (q, \overline q, g) \to h_1 + h_2 + \dots + h_n$. For a description of multiplicity distributions (MD) in this process, a two stage model (TSM) has been developed. The first stage of this process is described by pQCD \cite{NBD}, as the branching process that leads to the negative binomial distribution (NBD) of partons for a $q$-jet and to Pólya distribution for a $g$-jet. Hadronisation (the second stage) is described by the phenomenological scheme based on experimental data: at energies lower than 10 GeV where the hadronisation stage predominates, the second correlative moment ($f_2 = \overline {n(n-1)} - \overline {n}^2$) assumes negative values (with the increasing energy, $qg$-cascade is developed and $f_2$ changes sign from "$-$" to "$+$"). The binomial (Bernoulli) distribution is used for a description of the second stage. TSM's parameters of hadronisation have the following sense: the mean ($\overline n^h_g$) and max possible number of hadrons formed from $q$ or $g$ at their passing through the hadronisation stage. TSM is based on the convolution of these two stages and it describes MD in  $e^+e^-$ annihilation from 10 up to 200 GeV especially well in the high multiplicity (HM) region. Multiplicity in this region is considerably  more than the mean value: $n >> \overline n$. The gluon parameter $\overline n^h_g$ stays constant and close to 1 (Fig. \ref{fig:1}, left panel) in the whole investigating region (up to 200 GeV) confirming the fragmentation mechanism of hadronisation \cite{GDM}. This is also is known as hypothesis LocPHD.  TSM predicts the following values for the future $e^+e^-$ experiments at ILC and CLIC:  $\overline {n}_{ch}$~(500~GeV)~$\approx $~30$~\pm $~5,~$\overline {n}_{ch}$~(1~TeV)~$\approx $~50~$\pm $~10. 

\section{Hadronic and nuclear interactions}
\begin{figure}
\resizebox{\textwidth}{!}{%
  \includegraphics[width=0.4\textwidth]{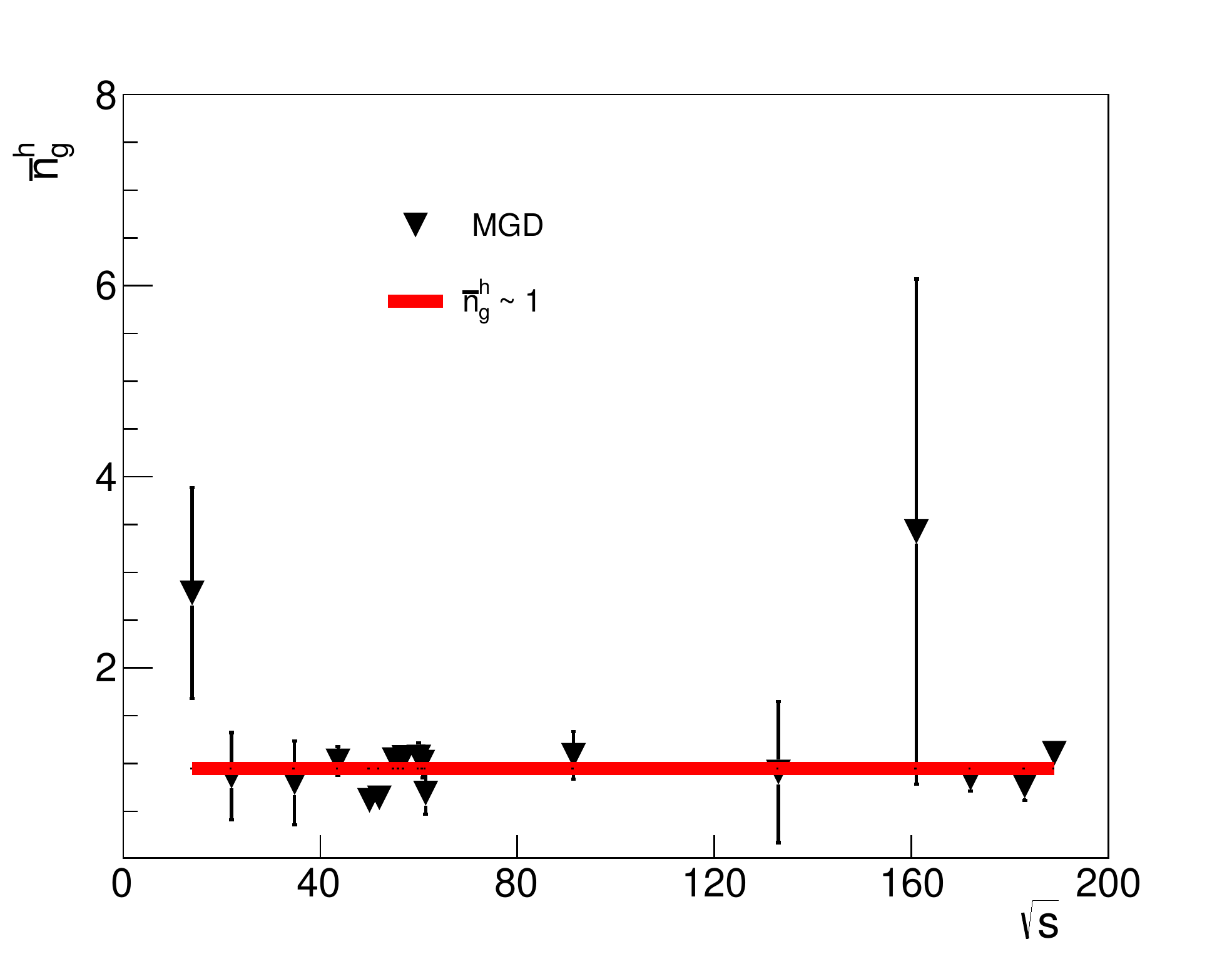}
  \includegraphics[width=0.4\textwidth]{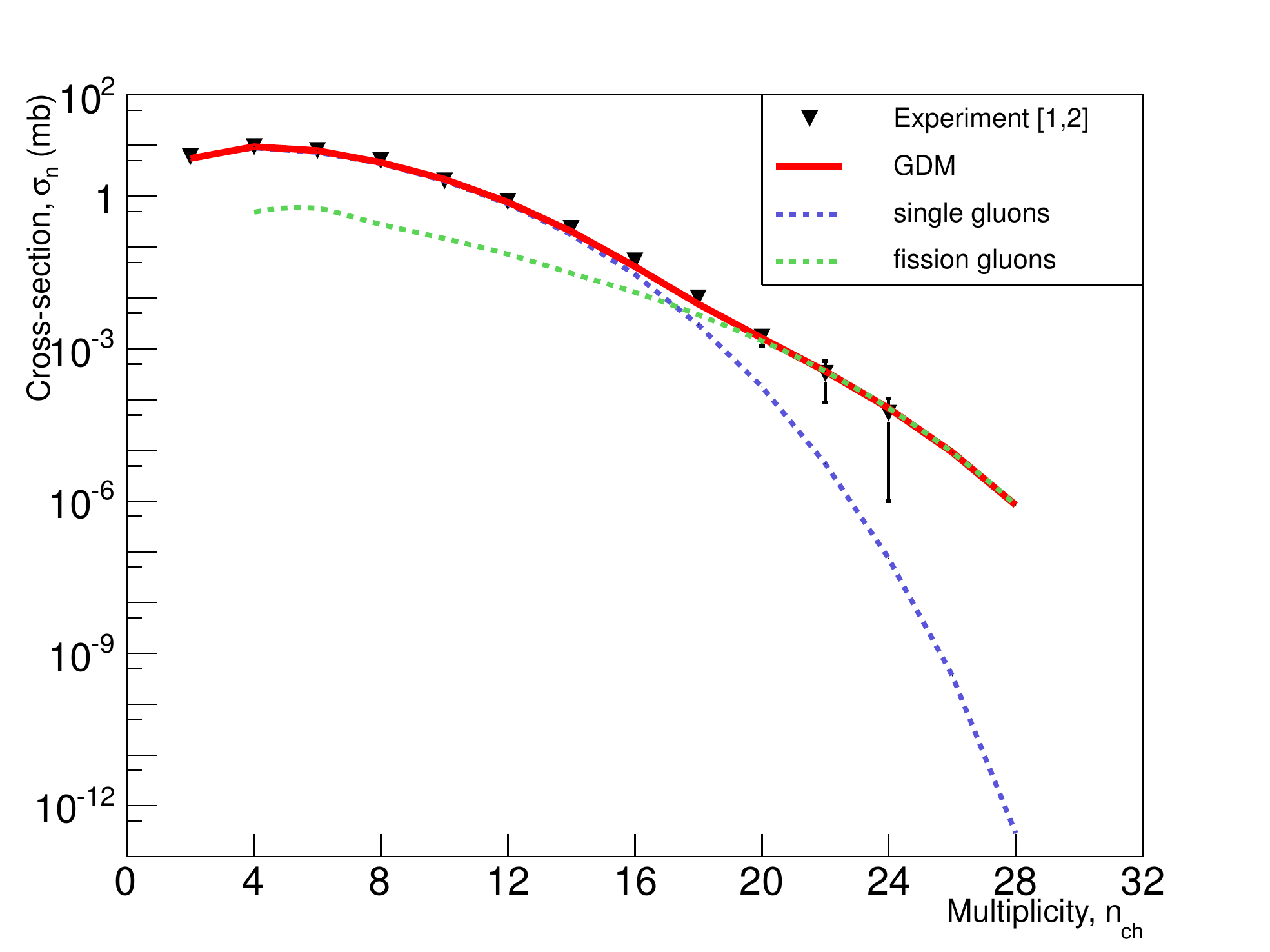}
}
\caption{
Left panel: Hadronization parameter of gluons $\overline n^h_g$ (\cite{GDM}). 
Experimental data using for their description by TSM are given in \cite{TSM}. Right panel:
Topological cross sections $\sigma _n$ versus charged multiplicity $n_{ch}$ in GDM. 
The dashed blue line describes the contribution of single sources, the green line 
$-$ the sources consisting of two gluons of fission, the solid red line is their sum.}
\label{fig:1}       
\end{figure}

\begin{figure}[]
\begin{minipage}[h]{0.4\linewidth}
\center{\includegraphics[width=1\linewidth]{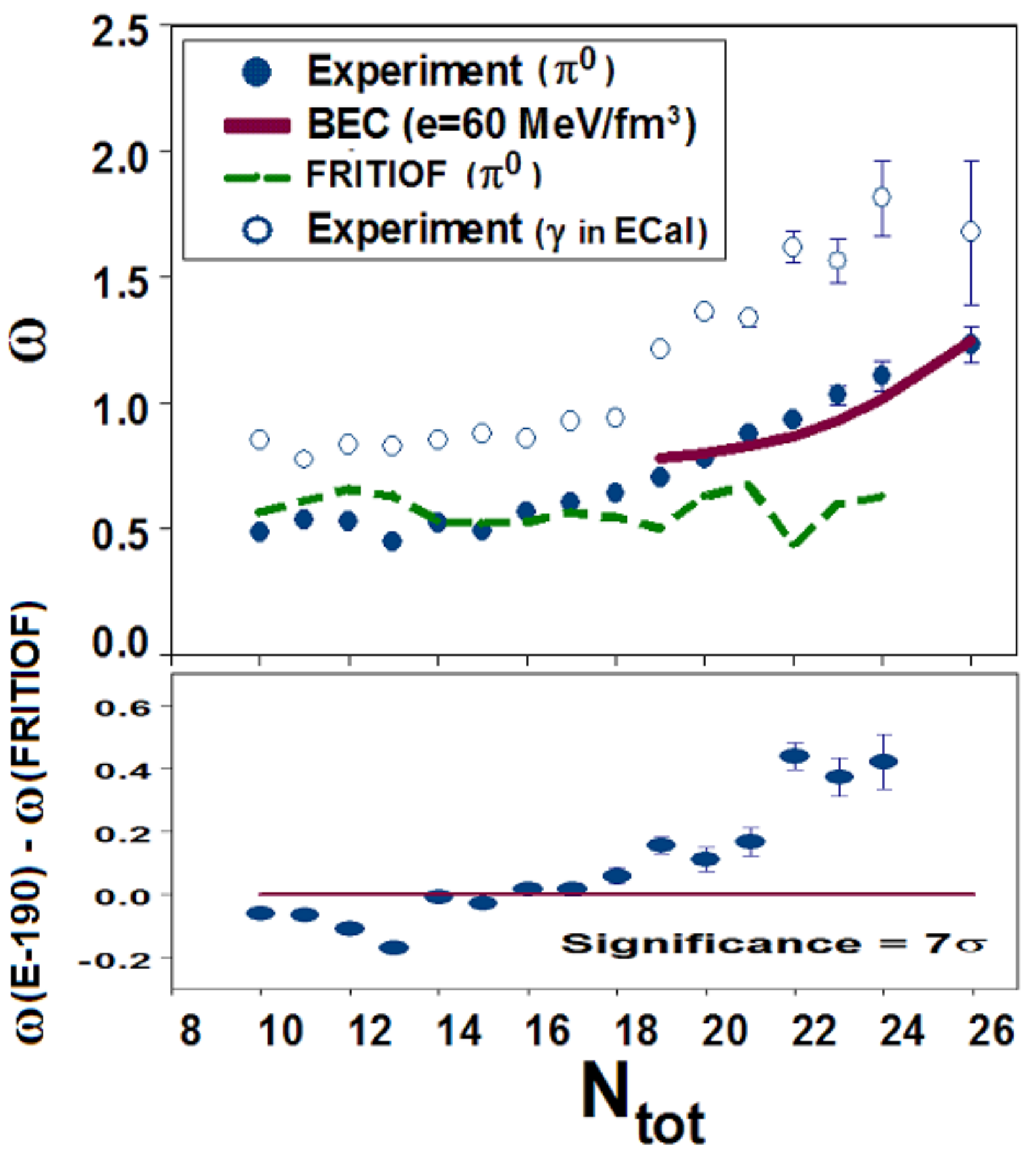}} a)\\
\end{minipage}
\hfill
\begin{minipage}[h]{0.45\linewidth}
\center{\includegraphics[width=1\linewidth]{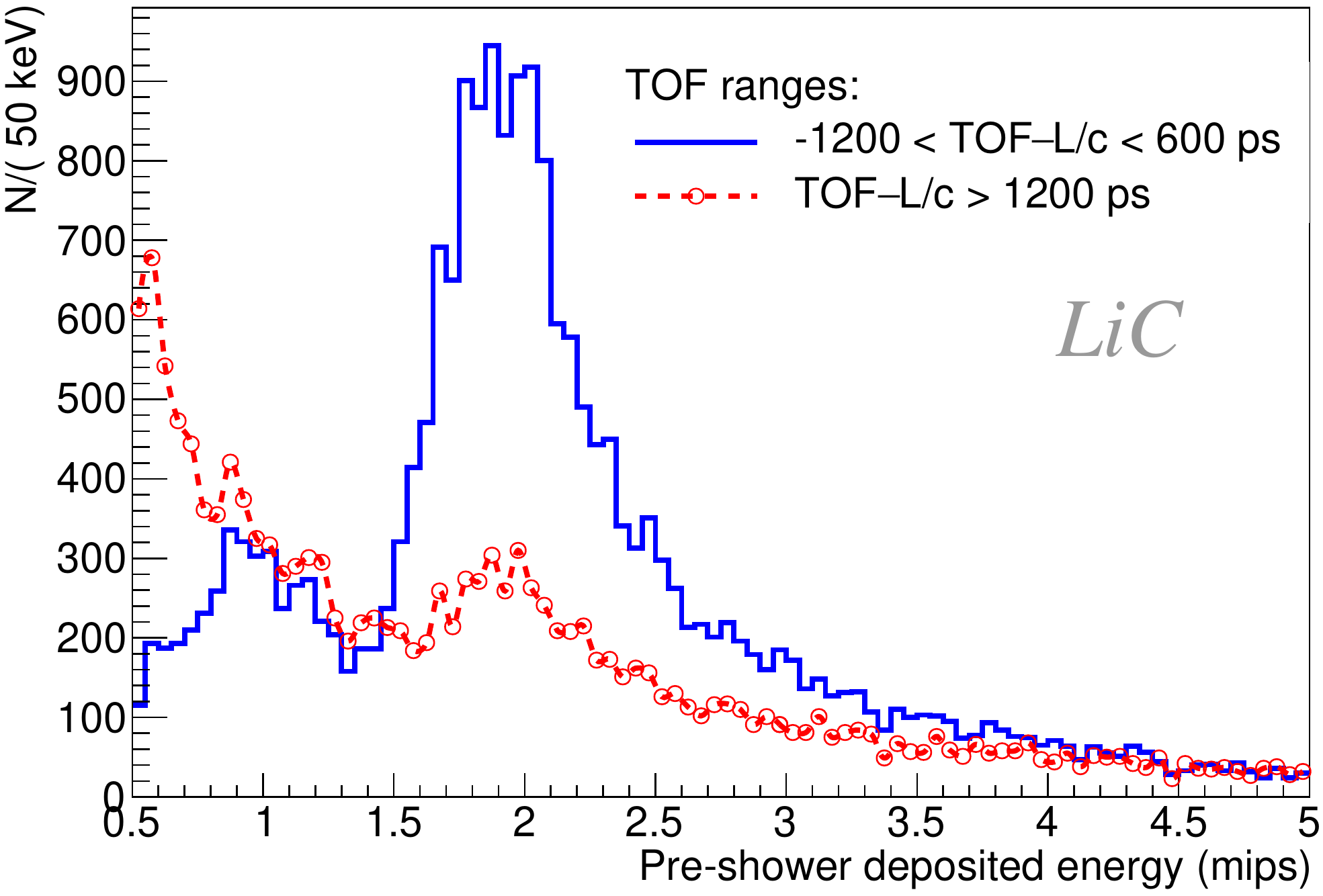}} b)\\
\end{minipage}
\vfill
\begin{minipage}[h]{0.5\linewidth}
\center{\includegraphics[width=1\linewidth]{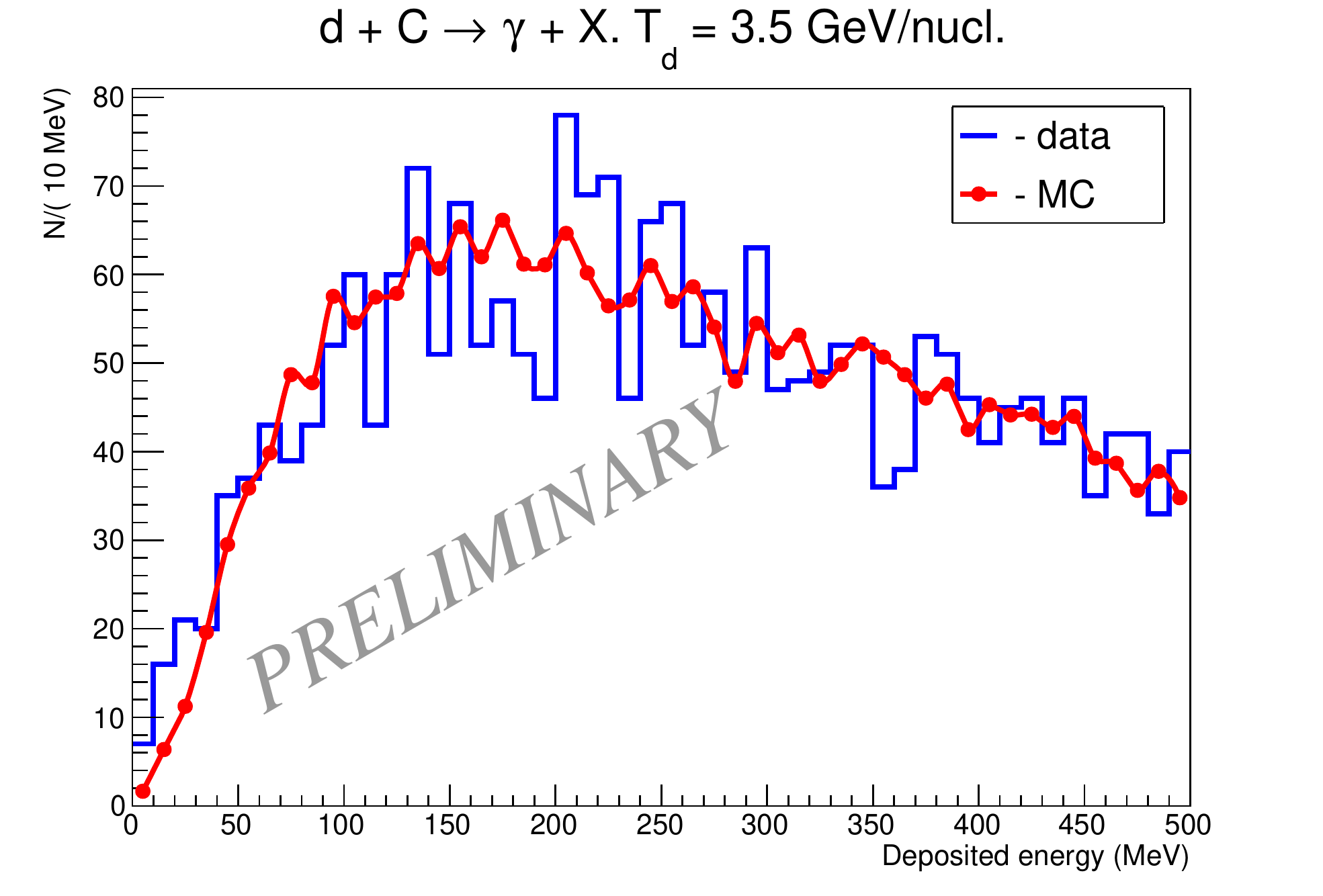}} c)\\
\end{minipage}
\begin{minipage}[h]{0.5\linewidth}
\center{\includegraphics[width=1\linewidth]{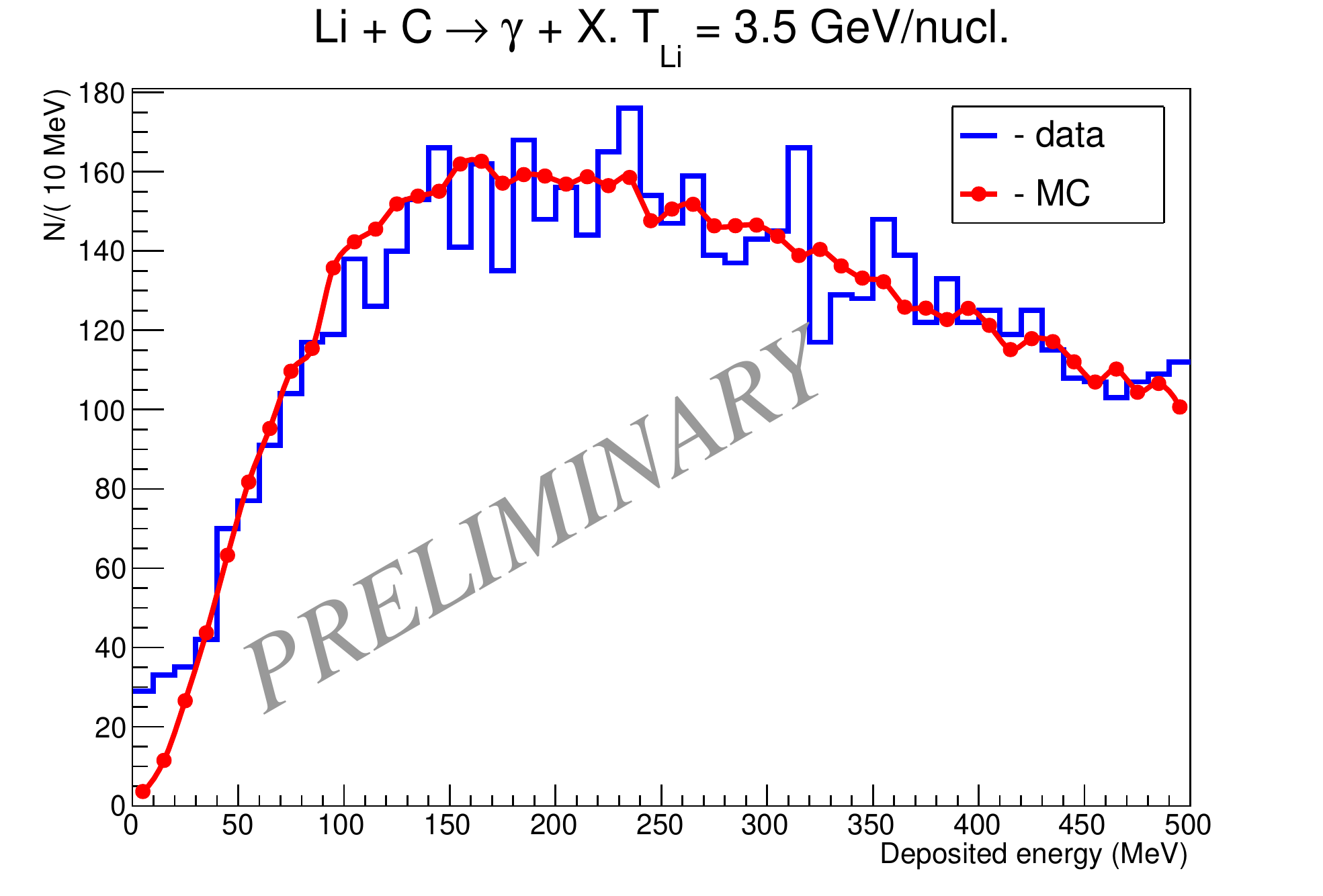}} d)\\
\end{minipage}
\caption{a) the measured scaled variance $\omega $ versus $N_{\mathrm {tot}}$ for $\pi ^0$-mesons
($\bullet $), photons ($\circ $), MC code FRITIOF7.02 (the dashed curve) and theoretical prediction (solid curve) \cite{Goren} for the
energy density $\varepsilon $ =60 MeV/fm$^3$. $N_{\mathrm {tot}} = N_{\mathrm {ch}} +N_0$ for $\pi ^0$-mesons and $N_{\mathrm {tot}} = N_\mathrm {{ch}} +N_{\gamma }$ for photons; b)
the distribution on the deposited energy in SPEC at different ToF of neutral particles
 for Li+C interactions, c) entire energy spectra in SPEC with pre-shower 
and simulation in d+C and
d) in Li+C interactions at Nuclotron.}
\label{fig:2}  
\end{figure}

Up to now, Monte Carlo simulation has difficulties in the description of MD in the HM region \cite{ATLAS}. They more often overestimate experimental topological cross sections. Phenomenological models give different predictions for them, too. That is why the SVD-2 Collaboration put the aim the investigating of this region in the proton interactions. This experiment was carried out at the U-70 accelerator (IHEP, Protvino) with a 50 GeV/$c$ proton beam on a H$_2$-target at the SVD-2 setup:  $p + p \to  2N + \pi _1 + \pi _2 + \dots + \pi _n$,  where $N$ $-$ nucleon, $n$ $-$ a pionic multiplicity. In the $pp$-interactions at the high energy, the multi-particle production occurs in $qg$-medium when valence quarks and lots of gluons can appear. Under QCD: valence quarks (gluons) can branch in accordance with such elementary processes as $q \to q + g$ and $g \to g + g$. To describe MD in the $pp$ interactions, we worked out a gluon dominance model (GDM). It is based on QCD and describes hadronisation by means of Bernoulli  as in TSM. The comparison of GDM with data is shown in Fig. \ref{fig:1}, right panel: valence quarks remain in the leading particles, and many gluons appear \cite{GDM}. But only a part of them (active gluons) is converted to hadrons, the rest gluons can be reradiated with the soft photons. GDM confirms the recombination mechanism of hadronization in the $qg$-medium for the  $pp$ and $p\overline p$ interactions (the hadronisation parameter of a gluon $\overline n^h_g$ exceeds of the corresponding value for the $e^+e^-$ annihilation). At the same time, the description of MD in the HM region can be improved by taking into account the fission of the active gluons \cite{kur}. Taking into account of the gluonic fission, we improve significantly the MD description at the HM region (Fig.~\ref{fig:1}, right).

In 2012, SVD-2 Collaboration has found the indication of the pionic (Bose-Einstein) condensate (BEC) formation in the $pp$-interactions in the region of the total (charged + neutral) HM multiplicity \cite{BEC}. The study of this phenomenon was begun in 70th \cite{Vosk}.  Over the years M.~Gorenstein and V.~Begun shown how to get evidence for BEC formation \cite{Goren}. We have ascertained: the growth of the scaled variance $\omega = D/\overline N_{0}$, $D = \overline {N_0^2} - \overline {N}_0^2$ with increasing of $N_{\mathrm {tot}}$ is observed at $N_{\mathrm {tot}} = N_{\mathrm {ch}} + N_0 \ge $ 18 and it gets 7 standard deviations relative to MC predictions \cite{BEC} as it is presented in Fig.~\ref{fig:2}, a). Then our Collaboration manufactured a Soft Photon Electromagnetic Calorimeter (SPEC) to verify the connection of the BEC formation and the anomalous yield of soft photons (energy < 50 MeV) \cite{EPJ}.

The SPEC has a low energy threshold of a registration ($E_{\mathrm {thresh}} \approx $ 1 MeV). 
The distribution of the deposited energy of neutral particles in the pre-shower in MIPs 
for Li+C interactions shown in Fig.~\ref{fig:2}, b)
for ToF smaller than 600 ps illustrates a Compton peak at 
1~MIP and more intensive peak for the gamma quanta 
conversion at 2 MIP. 
The spectra of the deposited energy obtained at Nuclotron (JINR) 
at interactions of  3.5A~ GeV/$\it c$  d and Li beams with a C-target 
reveal \cite{SP} the noticeable excess 
at energy less than 50 MeV in comparison with MC models. 
We plan to continue the soft photon study at Nuclotron andHM region.

\end{document}